\documentclass[prl,twocolumn]{revtex4-1}

\usepackage{graphicx}

\begin{document}

\title [Accelerated Bose-Einstein condensates in a double-well potential]{Accelerated Bose-Einstein condensates in a double-well potential}

\author {Andrea SACCHETTI}

\address {Department of Physics, Computer Sciences and Mathematics, University of Modena e Reggio Emilia, Modena, Italy}

\email {andrea.sacchetti@unimore.it}

\date {\today}

\thanks {This work is partially supported by Gruppo Nazione per la Fisica Matematica (GNFM-INdAM). \ The author is grateful for 
the hospitality of the Isaac Newton Institute for Mathematical Sciences where part of this paper was written.}

\begin {abstract} {Devices based on ultracold atoms moving in an accelerating optical lattice or double-well potential 
are a promising tool for precise measurements of fundamental physical constants as well as for the construction of sensors. \ Here, we 
carefully analyze the model of a couple of BECs separated by a barrier in an accelerated field and we show how the 
observable quantities, mainly the period of the beating motion or of the phase-shift, are related to the physical 
parameters of the model as well as to the energy of the initial state.}

\bigskip

{\it PACs numbers:} 03.75.Kk, 37.10.Jk, 03.75.Dg 

%

\end{abstract}

\maketitle

Laser-cooled atoms have 
drawn a lot of attention as for potential applications to interferometry and high-precision 
measurements, from the determination of gravitational constants to geophysical 
applications \cite {FFMK,LBCPT,MFFSK,RSCPT}, see also \cite {CladeReview,TinoReview} for a recent review. \ The idea of using cold atoms moving 
in an accelerating optical lattice \cite {Bloch1,Bloch2,RSN,SPSSKP,Shin} has open the field to multiple applications. \ For instance, by means of 
method proposed by Clad\'e {\it et al} \cite {CGSNJB} a precise measurement of the Earth's gravitational 
acceleration constant $g$ were performed \cite {PCC1,PCC2}; the 
obtained results had a very high precision and only a tiny discrepancy between $g$ measured by a Raman interferometry on laser-cooled atoms
and a classical gravimeter resulted, in fact the absolute relative uncertainty $\Delta g/g$ turns out to be of order 
$3 \times 10^{-9}$. 

More recently, a value for the constant $g$ has been measured using 
ultracold strontium 
atoms confined in an amplitude-modulated vertical optical lattice \cite  {PWTAPT}, improving a previous result 
\cite {FPST} by using a larger number of atoms and reducing the
initial temperature of the sample. \ Determination of $g$ 
has been obtained by measuring the frequency $\nu_B$ of the Bloch oscillations of the atoms in the 
vertical optical 
lattice and recalling that $\nu_B = m g d /2 \pi \hbar $, where $m$ is the mass of 
the Strontium atom, $\hbar$ is the Planck constant and $d$ is the lattice period. \ Since Bloch 
oscillations only occur for an one-body particle in a periodic field and under the effect of a Stark potential 
then has been chosen, in the experiment above, a particular Strontium's isotope ${}^{88}Sr$; in fact the 
scattering length $a_s$ of atoms ${}^{88}Sr$ is very small and thus it can be assumed that the effects of the atomic 
binary interactions are negligible. \ The obtained value for the constant $g$ 
was consistent with the previous one but was affected by a larger relative uncertainty of 
order $6 \times 10^{-6}$, because of a larger scattering in repeated measurements, mainly due to 
the initial position instability of the trap. \ Such a technique is also proposed to 
measure surface forces \cite {SAFIPST}.

On the other side, new technologies enable the construction of simple coherent matter-wave beam splitter 
based on atom chips. \ These devices have been shown to be capable of trapping and guiding ultracold 
atoms on a microscale; BECs can be efficiently created in such small devices and coherent quantum phenomena have 
been observed. \ Interferometers based on a microchip can be widely used as highly sensitive devices because they 
allow 
measurement of quantum phases. \ Technologically, chip-based atom interferometers promise to be very useful as 
inertial and 
gravitational field sensor 
provided that the quantum evolution of the matter waves is not perturbed by the splitting process. \ It has been seen 
that in such  a device a BEC cloud up to $10^5$ Rubidium-87 atoms can be split in two clouds inducing a 
double-well trapping potential phase-preserving \cite {SHAWGBSK}. \ By means of such a devices a measurement of the 
Earth's acceleration constant $g$ has been performed with relative uncertainty of 
order $2 \times 10^{-4}$ \cite {HWAHS}. \ It is well known that one of the most relevant physical effect in a 
double-well model is the so called \emph {beating motion} between the two wells; hence, in principle one can measure the beating period 
of the BEC in an accelerated double-well potential and then obtain the value of the gravitational constant, as done for BECs in 
an accelerated optical lattice. \ However, we would remark that in the case of ${}^{87}Rb$ isotopes the scattering length $a_s$ is not small 
and thus binary interactions must be effectively taken into account if one want to relate the Earth's gravitational constant $g$ 
with the beating motion of the two BEC's clouds between the two wells. \ Therefore, in order to improve the 
analysis of the experimental output it is necessary to have a more complete understanding of the underlying 
theory of two BECs separate by a asymmetrical barrier. 

The aim of this paper is to provide a solid theoretical ground for a BECs in a double well potentials under the 
effect of the gravity force, where an explicit formula connecting the physical 
parameters, and in particular the Earth's acceleration constant $g$, with the period of the observed beating motion 
between the two wells and of the difference of phase of two condensates. \ By means of such a result we expect that 
the relative uncertainty of the experimental results obtained for BECs in chips may be improved. \ Indeed, in such a framework the measure of the 
period of the difference of the phases between the two condensates gives a precise value for the Earth's acceleration 
constant $g$. \ We would underline that our analysis will be useful even as a model for a.c. Josephson effects in BECs 
\cite {LLSS} 

Here, we consider a simple model of BEC trapped in a double-well potential under the effect of a Stark potential, 
the dynamics along the direction of the gravity force is described by the one-dimensional Gross-Pitaevskii 
equation (GPE)
\begin{eqnarray}
\left\{
\begin {array}{l}
i \hbar \frac {\partial \psi }{\partial t} = - \frac {\hbar^2}{2m} \frac {\partial^2 \psi}{\partial x^2}+ V \psi + \epsilon  |\psi |^{2}  \psi + \nu x \psi  \\ 
\psi (x,0)=\psi_0 (x) 
\end {array}
\right. 
\, ,
\label {Equa1}
\end{eqnarray}
where $ V $ is the double-well trapping potential $V$, $\nu = m g$ is the strength of the Stark potential and the 
nonlinearity is given by $\epsilon = \frac {4 N \pi a_s \hbar^2}{m}$, with $N$ the total atom number, $a_s$ is the scattering length, 
$g$ is the gravity acceleration and $m$ is the atom mass. \ The BEC wavefunction $\psi$ is normalized to one.

By assuming the two-level approximation \cite {RSFS,OACM} then the normalized BEC wave function $\psi$ can be written as 
\begin{eqnarray*}
\psi (x,t) = e^{-i \Omega t /\hbar} \left [ a_R(t) \varphi_R (x) + a_L(t) \varphi_L (x) \right ] 
\end{eqnarray*}
where $a_{R,L} (t)$ are two complex valued functions depending on the time $t$ satisfying 
\begin{eqnarray*}
|a_R (t)|^2 + |a_L (t)|^2 = 1 ;
\end{eqnarray*}
the vector $\varphi_R$ (resp. $\varphi_L$) corresponds to the ground state of the corresponding isolated right hand side (resp. left hand side) 
trap with associated energy $\Omega$.

It is well known that the solution to the unperturbed problem (\ref {Equa1}), where $\epsilon =0$ and $\nu =0$ and 
when the state is initially prepared on the first two ground states, exhibits a beating motion with period $ \frac {\pi \hbar}
{\omega}$ independent of the initial wave function $\psi_0$, where $\omega = \frac {E_- - E_+}{2}$ is half of the the splitting between the two onsite 
energies $E_{\pm} = \Omega \mp \omega$. \ Hence, $\frac {\pi \hbar}
{\omega}$ plays the role 
of \emph {unit of time} and it is natural to introduce the (adimensional) \emph {slow time} 
\begin{eqnarray}
\tau = \frac {\omega t}{\hbar } \, . \label {tempo} 
\end{eqnarray}

The amplitudes $a_{R,L} (\tau )$ obey the nonlinear two-mode dynamical system given by 
(hereafter $' = \frac {d}{d\tau }$)
\begin{eqnarray}
\left \{
\begin {array}{lcl}
i a_R' &=& - a_L + \eta |a_R|^{2} a_R  + \rho  a_R \\ 
i a_L' &=& - a_R + \eta |a_L|^{2} a_L  - \rho  a_L 
\end {array}
\right. \, ,
\label {Equa8}
\end{eqnarray}
where $\eta$ and $\rho$ are the adimensional quantities defined as 
\begin{eqnarray*}
\eta = \frac {\epsilon }{\omega} \int_{-\infty}^{+\infty} |\varphi_R (x)|^4 dx 
\end{eqnarray*}
and
\begin{eqnarray*} 
\rho = \frac {\nu}{\omega} \int_{-\infty}^{+\infty} x |\varphi_R (x)|^2 dx \, . 
\end{eqnarray*}
The unperturbed solution (for $\eta = \rho =0$) of the two-level approximation has periodic solution with period $T = \pi$. 

If we set $a_{R,L} (\tau )= q_{R,L} (\tau )e^{i\theta_{R,L} (\tau )} $, where $q_{R,L} \in [0,1]$ are such that 
$q_R^2 (\tau )+ q_L^2 (\tau ) =1$, then the previous system (\ref {Equa8}) takes the Hamiltonian form
\begin{eqnarray}
\left \{ 
\begin {array}{lcl}
\theta ' &=&  -\frac {\partial {\mathcal H}}{\partial z} \\
z ' &=&  \frac {\partial {\mathcal H}}{\partial \theta}
\end {array}
\right. \label {Equa12}
\end{eqnarray}
where $\theta := \theta_R-\theta_L$ is the phase shift and $z:=q_R^2-q_L^2$ is the imbalance 
function between the two condensates, with Hamiltonian function
\begin{eqnarray}
{\mathcal H} = - 2 \sqrt {1-z^2} \cos \theta + \frac 12 \eta (1+z^2) + 2 \rho z \, . \label {Equa12Bis}
\end{eqnarray}

We should remark that (\ref {Equa12}) is invariant with respect to the change of the sign of $\epsilon$ and $\rho$; more precisely, if $\rho <0$ 
then we can switch to the case of $\rho >0$ by changing the signs $z\to -z$ and $\theta \to - \theta$. \ Similarly, if $\eta <0$ then 
we can switch to the case $\eta >0$ by $z\to -z$ and $\theta \to  \theta + \pi$.

It is a remarkable fact that equation (\ref {Equa12}) admits explicit periodical solutions, and that the period 
$T $ of the imbalance function 
$z(\tau )$, as well as of the phase shift $\theta (\tau )$, can be explicitly computed as function of the 
parameters $\eta$ and $\rho$ as well as of the initial wave function \cite {S}. \ Therefore, in principle, if one 
experimentally measure the 
inversion frequency then one can obtain a precise value for the acceleration constant $g$. \ In fact, let us 
denote by $E$ the energy 
value of the Hamiltonian ${\mathcal H}$ on the initial state: $E:= {\mathcal H} (z_0, \theta_0)$. \ Then, 
from (\ref {Equa12}) and (\ref {Equa12Bis}) we have that $z(\tau )$ is a solution to the following ordinary 
differential 
equation of first order:
\begin{eqnarray}
(z')^2=  a z^4+b z^3+cz^2+d z+ e \label {Equa13}
\end{eqnarray}
where we set $a = -\frac 14 \eta^2$, $b = -2 \eta \rho$, $c = E \eta - 4- \frac {\eta^2}{2} -  4 \rho^2 $, 
$d = 4 E \rho - 2 \eta \rho $ and  
$ e =E\eta  - E^2+4 - \frac 14 \eta^2 $. \ 
Equation (\ref {Equa13}) has solution given by means of the Weierstrass's elliptic function ${\mathcal P} (\tau ;g_2,g_3)$ with parameters
\begin{eqnarray*}
g_2 &:=& ae-\frac 14 bd+\frac {1}{12}c^2  \\ 
g_3 &:=& -\frac {1}{16} e b^2+\frac 16 e a c-\frac {1}{16} ad^2+\frac {1}{48}dbc-\frac {1}{216}c^3 
\end{eqnarray*}
The Weierstrass's elliptic function ${\mathcal P} (\tau ;g_2,g_3)$ is a doubly periodic function which 
real period coincides with the period $T$ of the phase shift and of the imbalance functions. \ In order to compute the 
real period $T$ let $e_j$, $j=1,2,3$, be the roots of the trinomial $4s^3 - g_2 s - g_3$; and 
let $\delta =g_2^3-27g_3^2$. \ If $\delta \ge 0$ then $e_j \in R$, $e_3<e_2 \le 0 < e_1$, and 
\begin{eqnarray*}
T = \frac {2 K(k)}{\sqrt {e_1-e_3}} \, , \ k = \frac {e_2-e_3}{e_1-e_3} 
\end{eqnarray*}
where $K$ denotes the complete elliptic integral defined as 
\begin{eqnarray*}
K( k ) = \int_0^1 \left [ (1-s^2) (1- k s^2 ) \right ]^{-1/2} ds \, . 
\end{eqnarray*}
On the other side, if $\delta < 0$ then $e_2 \in R$ and $e_3 = \bar e_1$, with $\Im e_1 \not= 0$, and 
\begin{eqnarray*}
T = \frac {2 K(k)}{\sqrt {H_2}}\, , \ k = \frac 12 - \frac {3e_2}{4H_2} \, ,\ H_2 = \sqrt {2 e_2^2+e_1 e_3} \, .
\end{eqnarray*}

In particular, when the nonlinear interaction is negligible, 
that is $\eta =0$, then the three solutions simply are $e_1 = \frac {2}{3} \left ( 1 + \rho^2 \right )$, 
$e_2 = e_3 = -\frac {1}{3} \left ( 1 + \rho^2 \right )$ and the period $T$ actually does not depend on the initial 
wave function $\psi_0$, but 
it only depends on $\rho$:
\begin{eqnarray}
T = \frac {\pi}{\sqrt {1+\rho^2}} \, . \label {etazero}
\end{eqnarray}
In particular, in the limit of $\rho =0$ we recover the unperturbed beating period $T = \pi$.

\begin{center}
\begin{figure}
\includegraphics[height=5cm,width=5cm]{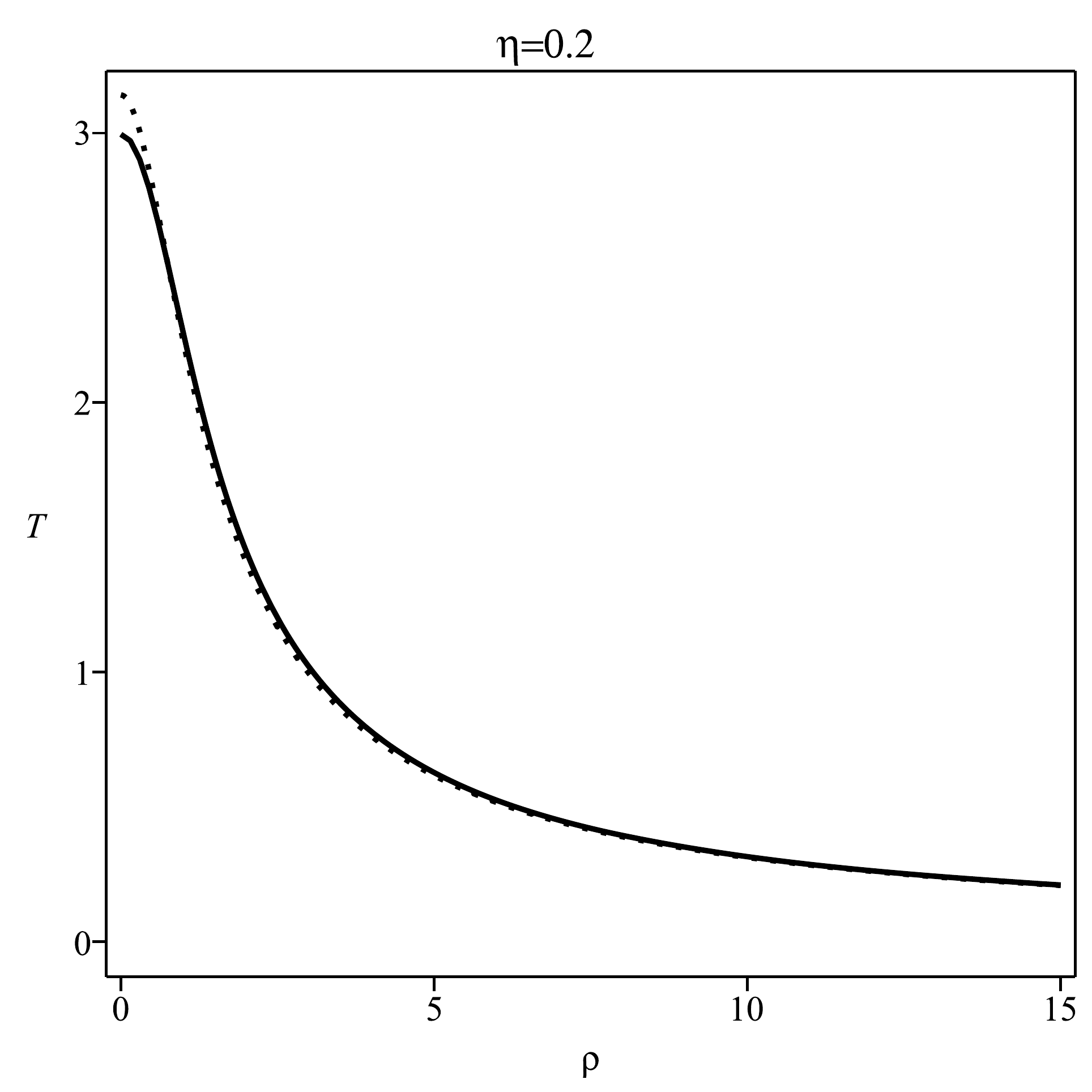}
\includegraphics[height=5cm,width=5cm]{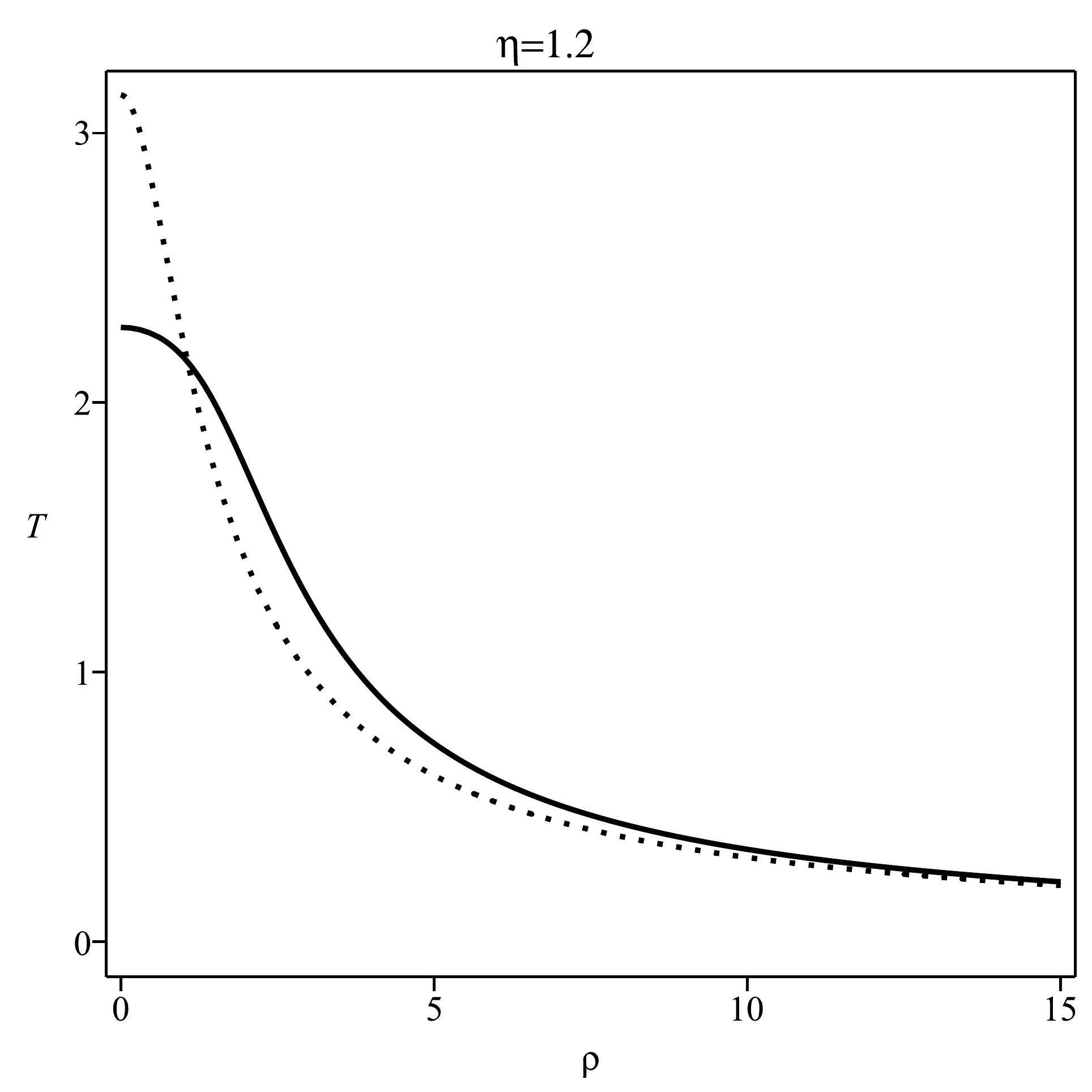}
\includegraphics[height=5cm,width=5cm]{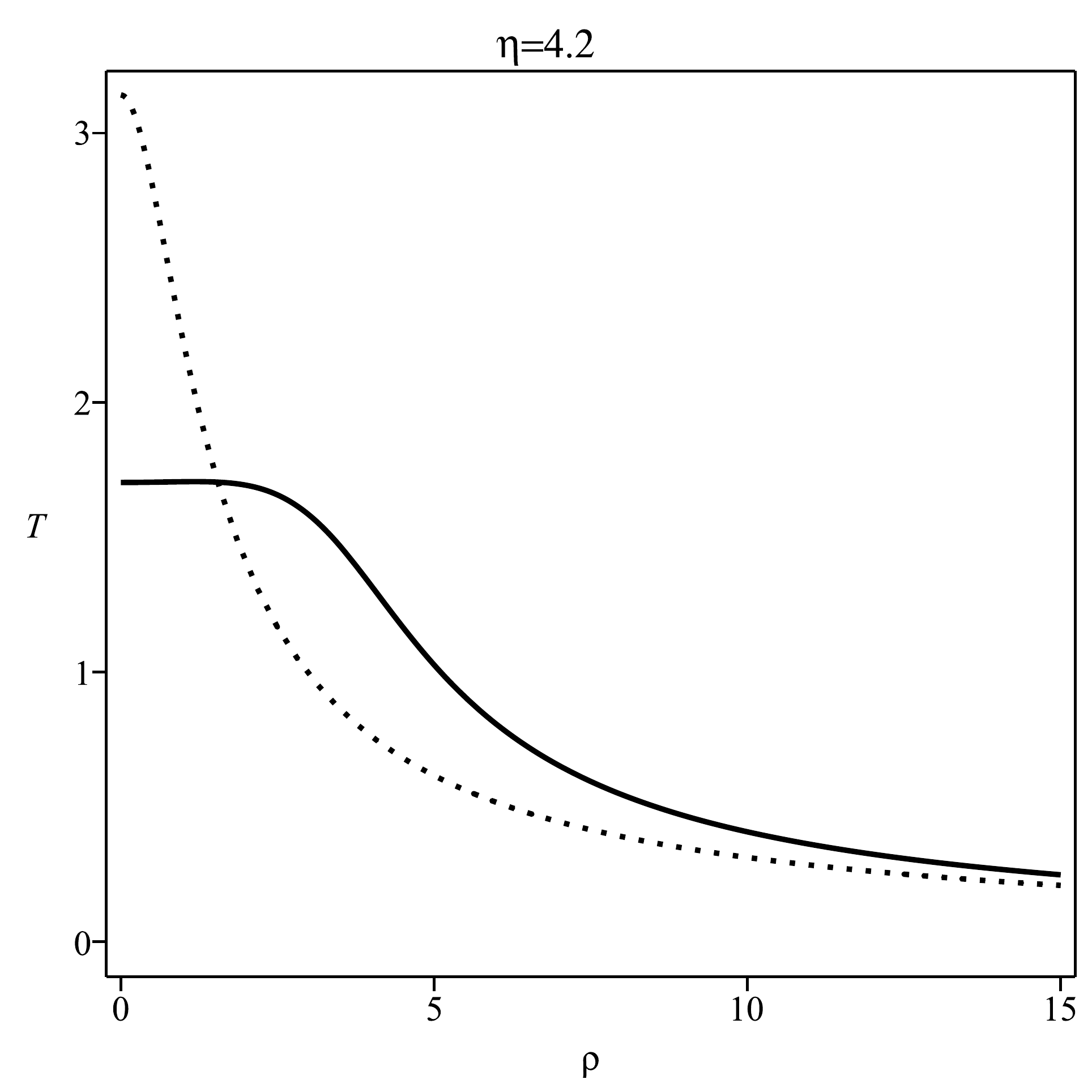}
\caption{\label {Fig1} In this figure we plot the 
graphs of the period $T$ versus the adimensional parameter $\rho$ for given values of $\eta =0.2$, 
$\eta =1.2$ and $\eta =4.2$ and when the initial stationary state $\psi_0$ is associated to 
a minimum value for the energy Hamiltonian ${\mathcal H}$ defined by (\ref {Equa12Bis}). \ Broken line 
correspond to the limit case of $\eta =0$, in such a case the period $T$ 
is given by equation (\ref {etazero}), and it does not depend on the initial wave function $\psi_0$}
\end{figure}
\end{center}

However, in general the period $T$ depends on the initial wave-function, as well as on the two parameters 
$\rho$ and $\eta$. \ In order to estimate the dependence of the beating period $T$ from the initial state and from 
the parameter $\eta$, corresponding to the strength of the nonlinear term, we consider, at first, the case where the 
initial wave function $\psi_0$ corresponds to a minimum value for the energy Hamiltonian ${\mathcal H}$ defined 
by (\ref {Equa12Bis}). \ As appears in Fig. \ref {Fig1} the period $T$, corresponding to the energy minimum, 
actually depends of $\eta$ for fixed value of $\rho$; for instance, the value of the period $T$ at $\rho =0$ 
and $\eta = 4.2$ is approximatively one half of the period $T$ at $\rho =0$ and $\eta =0$. \ Only for large 
value of $\rho$ we have a good agreement between the values of $T$ for different values of $\eta$.

\begin{center}
\begin{figure}
\includegraphics[height=5cm,width=5cm]{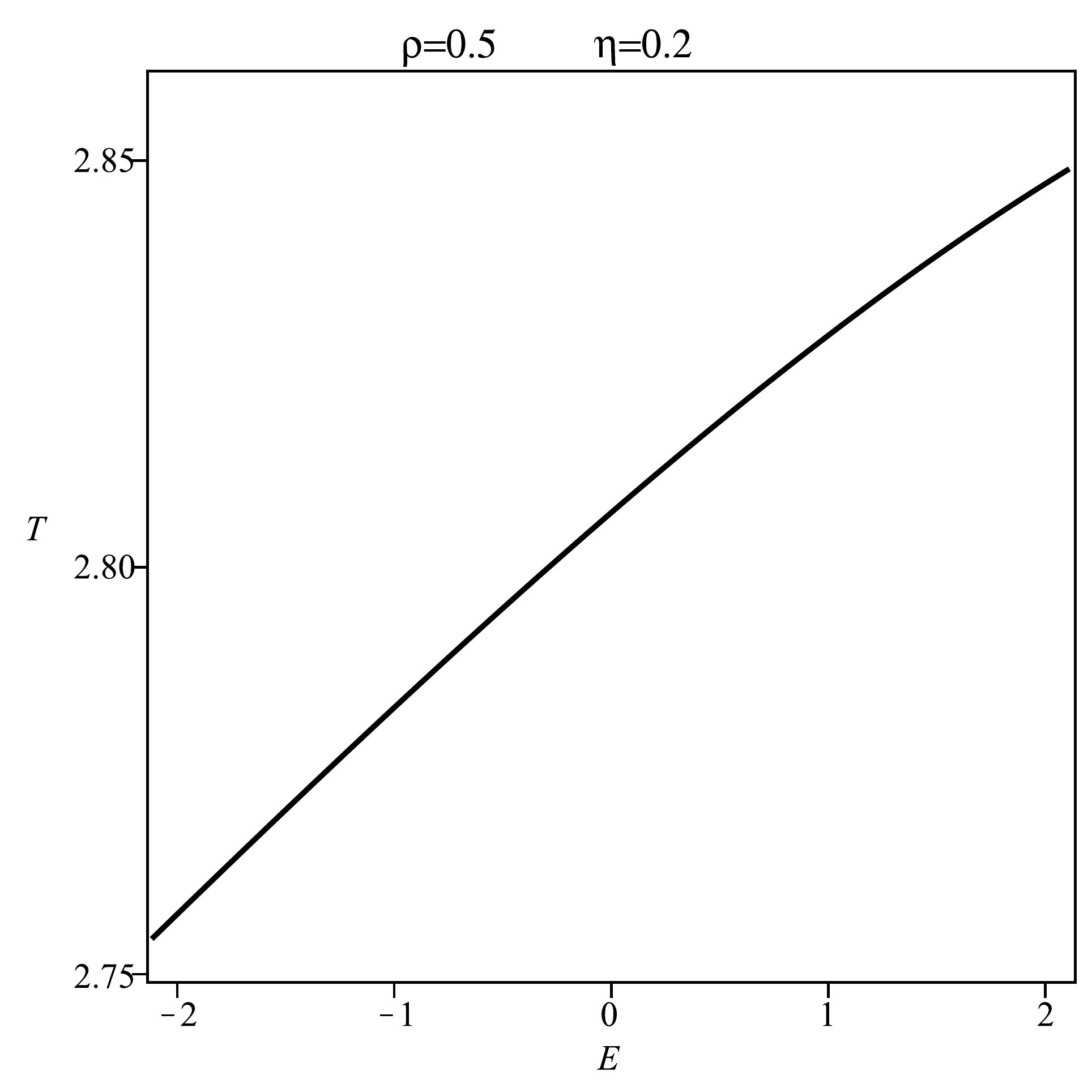}
\includegraphics[height=5cm,width=5cm]{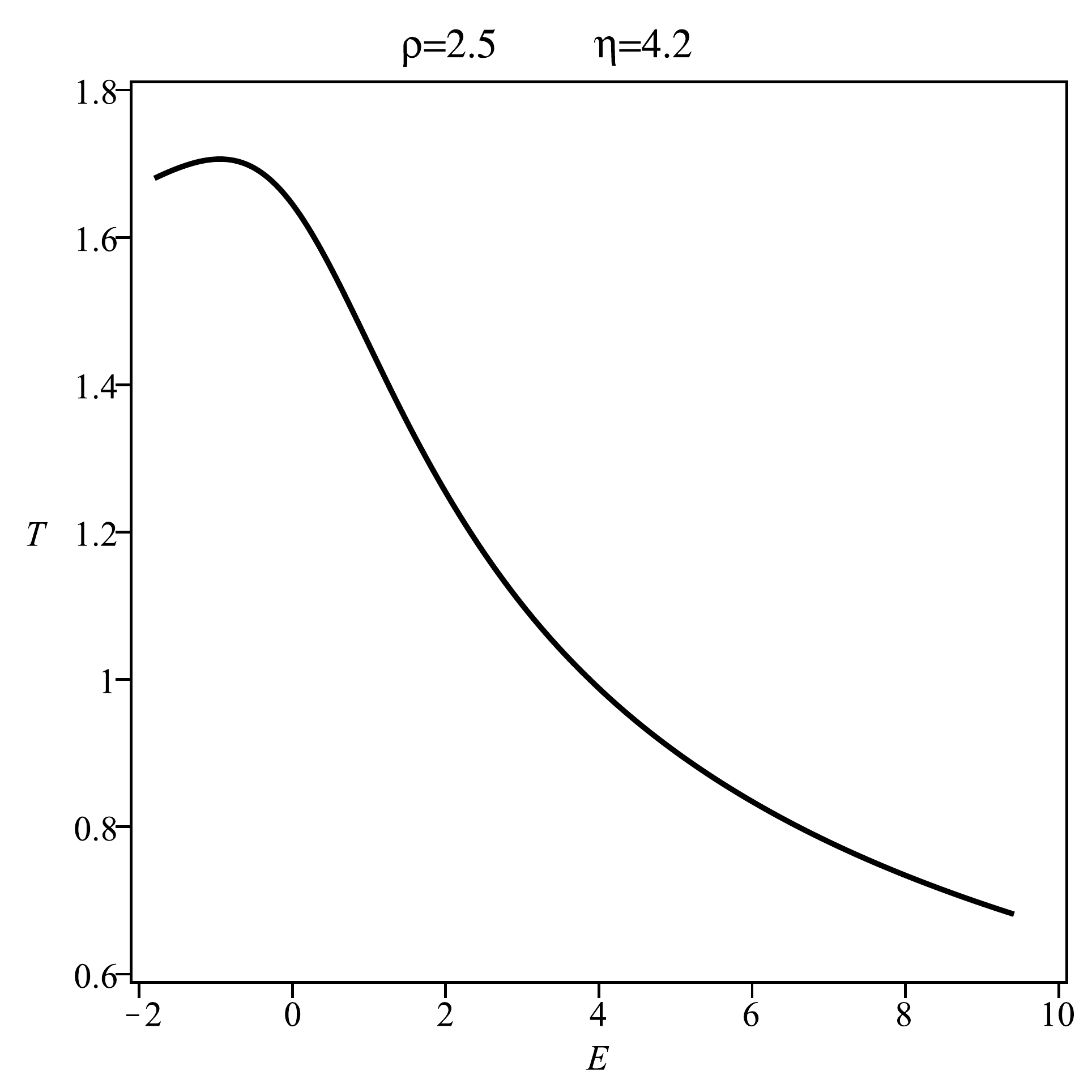}
\caption{\label {Fig2} In this figure we plot the graph of the period $T$ versus the values of the energy $E$ of 
the initial state for given values of $\rho $ and $\eta$.}
\end{figure}
\end{center}

We consider now the case where we fix the value of the adimensional parameters $\rho$ and $\eta$ and we 
compute the period $T$ as function of the energy of the initial state. \ For argument's sake we perform two 
numerical experiments (see Fig. \ref {Fig2}); in the first one we fix $\rho =0.5$ and $\eta =0.2$, while in the 
second one we fix $\rho = 2.5$ and $\eta =4.2$. \ For small values of the two parameters $\rho =0.5 $ and 
$\eta = 0.2$ then the period $T$ takes values (in the adimensional unit $\tau$ defined by (\ref {tempo})) from $T=2.754$ 
at $E=-2.117$ to $T=2.853$ at $E=2.358$, that is the period $T$ lies in an interval which length is around $1.7 \%$ 
of the mean value of $T$, the value of $T$ corresponding to the minimum value of 
the energy is $T=2.754$. \ On the other hand, for larger values of the two parameters $\rho =2.5 $ and 
$\eta = 4.2$ then the period $T$ takes values from $T=0.681$ at $E=9.416$ to $T=1.706$ at 
$E=-0.9586$, in such a case we have that the period $T$ lies in an interval which length is around $43 \% $ 
of the mean value of $T$, the value of $T$ corresponding to the minimum value of 
the energy is $T=1.68$. \ Hence, we can conclude that for some values of the parameters the period $T$ can 
strongly depend on the initial state.

We close by giving the expression of the adimensional quantities $\rho$ and $\eta$ as function of physical 
parameters in the semiclassical limit of small $\hbar$; these formulas will be useful in order to compute the period 
$T$ in a real device. \ To this end let us assume that the one-dimensional 
symmetric double-well potential is such that $V(-x) = V(x)$ with two non-degenerate absolute minima points at 
$x =\pm d$, where $2d$ is the distance between the bottom of the two wells, such that 
\begin{eqnarray*}
V (\pm d) <0 \, , \ \frac {dV (\pm d )}{dx} =0 \ \mbox { and } \  \mu :=\frac {d^2 V(\pm d)}{dx^2} >0 \, , 
\end{eqnarray*}
we assume also that the double-well potential goes to zero for large $x$. 
The eigenvalue equation $- \frac {\hbar^2}{2m} \frac {d^2 \varphi }{dx^2} + V \varphi = E \varphi $ admits two 
ground states $ E_\pm = \Omega \mp \omega$, where $\Omega = \frac 12 \hbar \sqrt {\mu/m} \left [ 1 + O (\hbar ) 
\right ] $ is the ground state energy of the single trap in the limit of small $\hbar$, with associated 
normalized eigenvectors $\varphi_{\pm}$. \ These eigenvectors are even and odd-parity functions 
$\varphi_\pm (-x) = \pm \varphi_\pm (x)$ and the normalized vectors $\varphi_{R,L}$ are given by 
\begin{eqnarray*}
\varphi_R = \frac {\varphi_+ + \varphi_- }{\sqrt {2}} \ \mbox { and } \ 
\varphi_L = \frac {\varphi_+ - \varphi_- }{\sqrt {2}}
\end{eqnarray*}
By construction, $\varphi_{R,L} (-x) = \varphi_{L,R} (x) $ and $\varphi_R$ is mostly localized on 
one of the two well (say the well with center at $x=+d$), and 
$\varphi_L$ is mostly localized on the other well (say the well with 
center at $x=-d$). \ In particular,
\begin{eqnarray*}
\varphi_R (x) = \varphi (x-d) \ \ \mbox { and } \ \ \varphi_L (x) = \varphi (x+d)
\end{eqnarray*}
where $\varphi (x)$ corresponds to the single trap eigenvector, that is 
\begin{eqnarray*}
\varphi (x) = a(x;\hbar ) e^{-\sqrt {m \mu } x^2 /2 \hbar} 
\end{eqnarray*}
with
\begin{eqnarray*}
a(x;\hbar ) = \frac {(m \mu )^{1/8}}{(\pi \hbar)^{1/4}} \left ( 1 + O(\hbar ) \right ) 
\end{eqnarray*}
in the semiclassical limit of small $\hbar$. \ Hence, if we denote by $a_H = \left [ \hbar^2 /m\mu \right ]^{1/4}$ 
the ground state oscillator length then the leading term of the parameters $\rho$ and $\eta$ are given by 
\begin{eqnarray*}
\rho = \frac {m g d}{\omega} \ \ \mbox { and } \ \ \eta = \frac {\epsilon }{\omega } 
\frac {1}{a_H \sqrt {2\pi }}
\end{eqnarray*}
where $\omega $ 
is half of the energy splitting, $2d$ is the distance between the bottoms of the two wells and 
$\epsilon$ is the strength of the Bose-Einstein condensate.

In conclusion: in this paper we have explored how relate the period $T$ of the beating motion and of the phase-shift 
of a couple of accelerated condensates separated by a barrier with the physical parameters of the nonlinear 
asymmetrical double-well model, mainly the strength of 
the nonlinear term, the strength of the Stark potential (which breaks the symmetry of the double-well potential) and 
the splitting between the first two onsite energies. \ We have also seen that in general the period may be strongly 
dependent of the energy of the initial state. \ Therefore, experimental determination of the physical constants 
(typically the Earth's acceleration constant $g$) should take into account such an effect.

\end {document}